\documentclass[%
 reprint,
 amsmath,amssymb,
 aps,
]{revtex4-2}

\usepackage{graphicx}% Include figure files
\usepackage{dcolumn}% Align table columns on decimal point
\usepackage{bm}% bold math
\usepackage{commath, mathtools,physics}
\usepackage{enumitem}
\usepackage{comment}
\usepackage{tabularx}
\usepackage{diagbox}
\usepackage{color}

\usepackage[caption=false]{subfig}
\usepackage{hyperref}% add hypertext capabilities
\usepackage{commands}

\begin{document}

% \preprint{APS/123-QED}

\title{Unified simulation methods for quantum acoustic devices}

\author{Hugo Banderier}
\altaffiliation[Current affiliation : ]{Oeschger Centre for Climate Change Research and Institute of Geography, Universit{\"a}t Bern, 3012 Bern, Switzerland}%
\affiliation{Department of Physics, Eidgen{\"o}ssiche Technische Hochschule Z{\"u}rich, 8093 Z{\"u}rich, Switzerland
}%
\affiliation{Quantum Center, Eidgen{\"o}ssiche Technische Hochschule  Z{\"u}rich, 8093 Z{\"u}rich, Switzerland
}%
\author{Maxwell Drimmer}
\altaffiliation[Corresponding author : ]{max.drimmer@phys.ethz.ch}
\affiliation{Department of Physics, Eidgen{\"o}ssiche Technische Hochschule  Z{\"u}rich, 8093 Z{\"u}rich, Switzerland
}%
\affiliation{Quantum Center, Eidgen{\"o}ssiche Technische Hochschule  Z{\"u}rich, 8093 Z{\"u}rich, Switzerland
}%
\author{Yiwen Chu}
\affiliation{Department of Physics, Eidgen{\"o}ssiche Technische Hochschule  Z{\"u}rich, 8093 Z{\"u}rich, Switzerland
}%
\affiliation{Quantum Center, Eidgen{\"o}ssiche Technische Hochschule  Z{\"u}rich, 8093 Z{\"u}rich, Switzerland
}%

\date{\today}

\begin{abstract}
In circuit quantum acoustodynamics (cQAD), superconducting circuits are combined with acoustic resonators to create and control non-classical states of mechanical motion. Simulating these systems is challenging due to the extreme difference in scale between the microwave and mechanical wavelengths. All existing techniques simulate the electromagnetic and mechanical subsystems separately. However, this approach may not be adequate for all cQAD devices. Here, we demonstrate a single simulation of a superconducting qubit coupled to an acoustic and a microwave resonator and introduce two methods for using this simulation to predict the frequencies, coupling rates, and energy-participation ratios of the electromechanical modes of the hybrid system. We also discuss how these methods can be used to investigate important dissipation channels and quantify the nontrivial effects of mode hybridization in our device. Our methodology is flexible and can be extended to other acoustic resonators and quantum degrees of freedom, providing a valuable new tool for designing hybrid quantum systems.

\end{abstract}

\maketitle

%\tableofcontents
\section{Introduction}
Circuit quantum acoustodynamics (cQAD) provides the opportunity to combine the unique advantages of superconducting (SC) circuits and mechanical resonators~\cite{Chu2020,clerk2020hybrid}. A device that integrates the numerous long-lived modes of compact mechanical resonators~\cite{MacCabe2020, Gokhale_epiHBAR_2020, Tsaturyan2017} with the strong quantum nonlinearity of SC qubits is potentially useful for quantum information processing~\cite{Hann2019,Pechal2019, chamberland2022building} and tests of fundamental physics~\cite{Pikovski2012}. SC qubits have already been combined with a wide variety of mechanical elements, including membranes~\cite{pirkkalainen2013hybrid, viennot2018phonon}, bulk acoustic wave (BAW) resonators~\cite{OConnell2010, ChuScience2017, kervinen2019landau}, surface acoustic wave resonators~\cite{Gustafsson2014, moores2018cavity, Satzinger2018}, and phononic crystals~\cite{Arrangoiz-Arriola2018}. 

Finite element (FE) simulations are a crucial design tool in both circuit quantum electrodynamics (cQED)~\cite{Nigg2012, Solgun2014BBQ, Solgun2015, Solgun2015multimode, Minev2021a, Minev2021} and solid mechanics~\cite{comsol_sm} for predicting the behavior of complex solid-state structures. Both fields have independently developed mature techniques which rely on different strategies and software. For example, quantum circuits are modeled using electromagnetic simulation software like Ansys HFSS~\cite{Ansys}, Microwave Office~\cite{MicrowaveOffice}, or Sonnet~\cite{Sonnet} while \textsf{COMSOL Multiphysics} (COMSOL)~\cite{COMSOL} is the preferred choice for performing FE simulations of acoustic resonators. 

Techniques from cQED and solid mechanics can be combined in order to simulate cQAD devices. If an electromechanical system can be modeled as a lumped element, it is possible to simulate the electromechanical response in isolation from the Josephson circuit~\cite{Arrangoiz-Arriola2016, Gely2020}. This response is represented by an equivalent circuit which can be quantized to determine the system's Hamiltonian~\cite{devoret1997fluctuations}. In certain cases, however, isolating an electrically small subsystem may not be possible meaning that the system cannot be described by simple circuit. The $\hbar$BAR~\cite{Chu2018, vonLuepke2021parity}, a cQAD device that features a 3-D transmon qubit~\cite{Paik2011} piezoelectrically coupled to a high-overtone BAW resonator (HBAR), falls into this category.

In this paper, we demonstrate two simulation approaches that unify FE techniques from cQED and solid mechanics using a single COMSOL model without the need for an equivalent circuit. In the first approach, which we call the ``unhybridized eigenmode approach," we solve for the eigenmodes of the electric and displacement fields separately (i.e. without any piezoelectric coupling). This yields the mode structure of the unhybridized electrical and mechanical subsystems. One can then evaluate the electromechanical coupling rate between an electrical and a mechanical eigenmode using an overlap integral inside the piezoelectric material (Eq.~\ref{eq:couplingRates}). In the second approach, which we call the ``hybridized eigenmode approach", we simultaneously solve for the coupled electric and displacement fields to find the dressed eigenmodes of the entire system. We demonstrate this approach by first presenting a Hamiltonian formulation that extends the energy-participation ratio (EPR) method~\cite{Minev2021a} to mechanical degrees of freedom. We then use it to extract important Hamiltonian parameters that arise from electromechanical coupling, such as cross-Kerr nonlinearities, anharmonicities, and mechanical EPRs in the dispersive regime. While simulating the coupled fields is more computationally intensive, the results reflect that hybridization of the electromagnetic and mechanical subsystems affects the frequency and shape of the modes, which can in turn impact the predicted coupling and loss rates. 

\section{Hybrid quantum interactions} \label{sec:Interactions}
We consider a general system consisting of a SC circuit with $J$ transmons and $N-J$ linear electromagnetic modes interacting with a mechanical resonator supporting $M$ modes. Even though the $N$ linear electromagnetic modes and $M$ mechanical modes are identically described as modes of a bosonic resonator, we use different letters to distinguish them for clarity in the rest of this section. Each of the $J$ transmons can be described using a Hamiltonian that is a sum of a linear resonator term and a nonlinear term~\cite{devoret1997fluctuations}. When the interactions are written under the rotating wave approximation, the Hamiltonian is
\begin{multline} \label{eq:H0}
    \frac{\hat{H}}{\hbar}= \sum_n^N \Tilde{\omega}_n \hat{\tilde{a}}_n^\dagger \hat{\tilde{a}}_n + \sum_m^M \Omega_m \bdag_m \bhat_m + \sum_{n,m}^{N,M} \Tilde{g}_{nm} \left(\hat{\tilde{a}}_n^\dagger \bhat_m + \bdag_m \hat{\tilde{a}}_j \right) \\ - \sum_{n\neq n'}^{N,N} \varsigma_{nn'} \left(\hat{\tilde{a}}_n^\dagger \hat{\tilde{a}}_{n'} + \hat{\tilde{a}}_{n'}^\dagger \hat{\tilde{a}}_n \right)  - \sum_j^J \frac{E_j}{\hbar} \left(\cos \hat{\theta}_j + \frac12 \hat{\theta}_j^2 \right) 
\end{multline}
where we have introduced $\tilde{\omega}_n$ and $\hat{\tilde{a}}_{n}$ ($\Omega_m$ and $\hat{b}_{m}$) as the frequencies and bosonic ladder operators of the $n$th ($m$th) linear electromagnetic (mechanical) resonator mode, the Josephson energy of the $j$th junction $E_j$ and its flux $\hat{\theta}_j = \theta_{\text{ZPF}, j}\left(\hat{\tilde{a}}_j + \hat{\tilde{a}}_j^\dagger\right)$ where $\theta_{\text{ZPF}, j}$ are the associated zero-point fluctuations (ZPFs), and $\varsigma_{nn'}$ ($\tilde{g}_{nm}$) are the electromagnetic (electromechanical) two-mode coupling rates. The detuning between two modes are defined as $\Delta_{nn'} = \tilde{\omega}_{n'} - \tilde{\omega}_n$ and $\Delta_{nm} = \Omega_m-\tilde{\omega}_n$. The coupling between two modes are said to be \textit{dispersive} if $|\Delta_{nn'}| \gg |\varsigma_{nn'}|$ or $|\Delta_{nm}| \gg |g_{nm}|$.

Eq.~\ref{eq:H0} is a nonlinear Hamiltonian that cannot be directly modeled using standard FE simulation techniques. Instead, we follow Ref.~\cite{Nigg2012} by rewriting the Hamiltonian as a sum of linear and nonlinear terms. Then we can use FE simulations to find the linear eigenmodes, from which we extract the relevant parameters that describe the nonlinear terms using the EPR method.

\subsection{Unhybridized eigenmode approach}\label{sec:UnhybridizedApproach}
If we ignore the coupling between the electromagnetic and mechanical degrees of freedom (third term in Eq.~\ref{eq:H0}), we can simulate the two subsystems individually. The results of electromagnetics-only simulations (i.e. simulations with solid mechanics and piezoelectricity turned off) are dressed eigenstates of the linear electromagnetic part of the Hamiltonian (Terms 1 and 4 of Eq.~\ref{eq:H0}). Thus the Hamiltonian can be partially diagonalized
\begin{multline} \label{eq:EMBB}
    \frac{\hat{H}}{\hbar}=\sum_{n=1}^N \omega_n \adag_n\ahat_n + \sum_m^M \Omega_m \bdag_m \bhat_m  + \sum_{nm}^{N,M} g_{nm}\left(\adag_n\bhat_m + \bdag_m\ahat_n \right) \\ - \sum_j^J \frac{E_j}{\hbar} \sum_{p=4}^\infty c_{p} \left(\sum_{n=1}^N \varphi_{nj} \hat{a}_n + \text{H.c.}\right)^p
\end{multline}
In this expression, $\hat{a}$ are the dressed electromagnetic ladder operators, $c_p$ are the cosine expansion coefficients, and $\varphi_{nj}$ are the ZPFs of the flux in the $j$-th junction when only dressed mode $n$ is excited and $E_j$ are the $j$-th junction's Josephson energies.

The solutions of the mechanical simulations are exactly the modes described Term 2 of Eq.~\ref{eq:H0}. The electromechanical interaction is now written in terms of the dressed electromagnetic eigenmodes. The exact expression of $g_{nm}$ depends on the nature of the interaction. Here, we will consider the piezoelectric coupling which we describe in detail in Appendix~\ref{sec:piezoHam}. From now on we will refer to the dressed modes $\hat{a}_n, \hat{b}_m$ of this picture as ``unhybridized" because the next approach will further hybridize these.

\subsection{Hybridized eigenmode approach}\label{sec:HybridizedApproach}
As there is no conceptual difference between an electromagnetic and a mechanical mode in Eq.~\ref{eq:H0}, we can equivalently choose to express the Hamiltonian in terms of $N+M$ hybrid electromechanical modes with eigenvalues $\xi_k$, bosonic ladder operators $\hat{c}_k$, and associated junction flux ZPFs $\phi_{kj}$

\begin{equation} \label{eq:BBcoupled}
    \frac{\hat{H}}{\hbar}=\sum_{k=1}^{N+M} \xi_k \hat{c}^\dagger_k\hat{c}_k - \sum_j^J \frac{E_j}{\hbar}  \sum_{p=4}^\infty c_{p} \left(\sum_{k=1}^{N+M} \phi_{kj} \hat{c}_k + \text{H.c.}\right)^p
\end{equation}
Under the dispersive and the perturbative assumptions, detailed in Appendix~\ref{sec:AppDisp}, we can limit the expansion of the second term to $p=4$ and only keep only excitation number-preserving interactions to obtain
\begin{equation} \label{eq:Hdisp}
    \frac{\hat{H}_{p=4}}{\hbar} = \sum_k -\Delta_k \hat{c}^\dagger_k \hat{c}_k - \frac12 \alpha_k \hat{c}^{\dagger 2}_k \hat{c}^2_k -
    \sum_{l<k} \frac12 \chi_{kl}\hat{c}^\dagger_k \hat{c}_k \hat{c}^\dagger_l \hat{c}_l
\end{equation}
This expression highlights several experimentally relevant quantities: $\Delta_k = \frac12 \sum_l \chi_{kl}$ are the effective Lamb shifts, $\alpha_k$ are the anharmonicities, and $\chi_{kl} = \hbar^{-1}\sum_j E_j \phi_{kj}^2\phi_{lj}^2 $ are the total cross-Kerr shifts induced between modes $k$ and $l$. Under these approximations all the parameters depend on the zero-point fluctuations of the junctions' fluxes in each mode, $\phi_{kj}$. We note that in cQAD, the perturbative assumption is not always valid in the dispersive regime. However, as shown in Appendix~\ref{sec:AppDisp}, the correction to these quantities can still be expressed using the same ZPFs. 

In the two approaches described above, we have developed Hamiltonians with key unknowns that can be obtained from simulations. In the unhybridized eigenmode approach, they are the bare mode frequencies $\omega_n$ and $ \Omega_m$, the junction flux ZPFs in the bare electromagnetic modes $\varphi_{nj}$, as well as the piezoelectric pairwise coupling rates $g_{nm}$. In the hybridized eigenmode approach, we only need to solve for the hybridized mode frequencies $\xi_k$ and the junction flux ZPFs $\phi_{kj}$.

\section{Simulating hybrid quantum devices} \label{sec:Simulations}

\subsection{Physics interfaces} \label{sec:modeling}

We begin each simulation by choosing physics interfaces in COMSOL, each of which defines a vector field along with its equations of motion to be solved. We used the \textbf{Electromagnetic Waves, Frequency Domain} interface (\textbf{emw}) in the \textsf{RF Module} for modeling SC circuits and microwave cavities and used the \textbf{Solid Mechanics} interface (\textbf{solid}) in the \textsf{Structural Mechanics Module} for the acoustic resonator. 

The second step is dynamically combining these interfaces together. In general, linking interfaces A and B is simply done by calling the field defined in A in a domain or boundary condition on B and vice versa. In many cases, COMSOL has built-in multiphysics interfaces that perform this step. However, no such interface exists between \textbf{solid} and \textbf{emw}. Therefore, the coupling must be defined manually. In this work, electromagnetic and mechanical objects are coupled by the piezoelectric effect \footnote{A piezoelectric multiphysics interface exists, but can only couple solid mechanics to the electrostatics interface, which is not suitable for simulating cQED devices. The electrostatics interface is unable to simulate a microwave cavity as it lacks a feature for phase propagation, and does not have a lumped element boundary condition.}. In a piezoelectric medium, the wave equations for the electric and displacement fields are modified to become Eqs.~\ref{eq:EOM_PZ1} and~\ref{eq:EOM_PZ2} as derived in Appendix~\ref{sec:piezoHam}. These modifications are implemented in our simulation using three additional domain conditions, represented as \textit{nodes} of the physics interfaces. Specifically, the \textbf{effective medium} and \textbf{external current density} nodes are added to the \textbf{emw} interface and an \textbf{external stress} node is added to the \textbf{solid} interface. A detailed description is provided in Appendix~\ref{sec:Setup}.  

\subsection{Model of the device}
\begin{figure}[t]
\includegraphics[width=\linewidth]{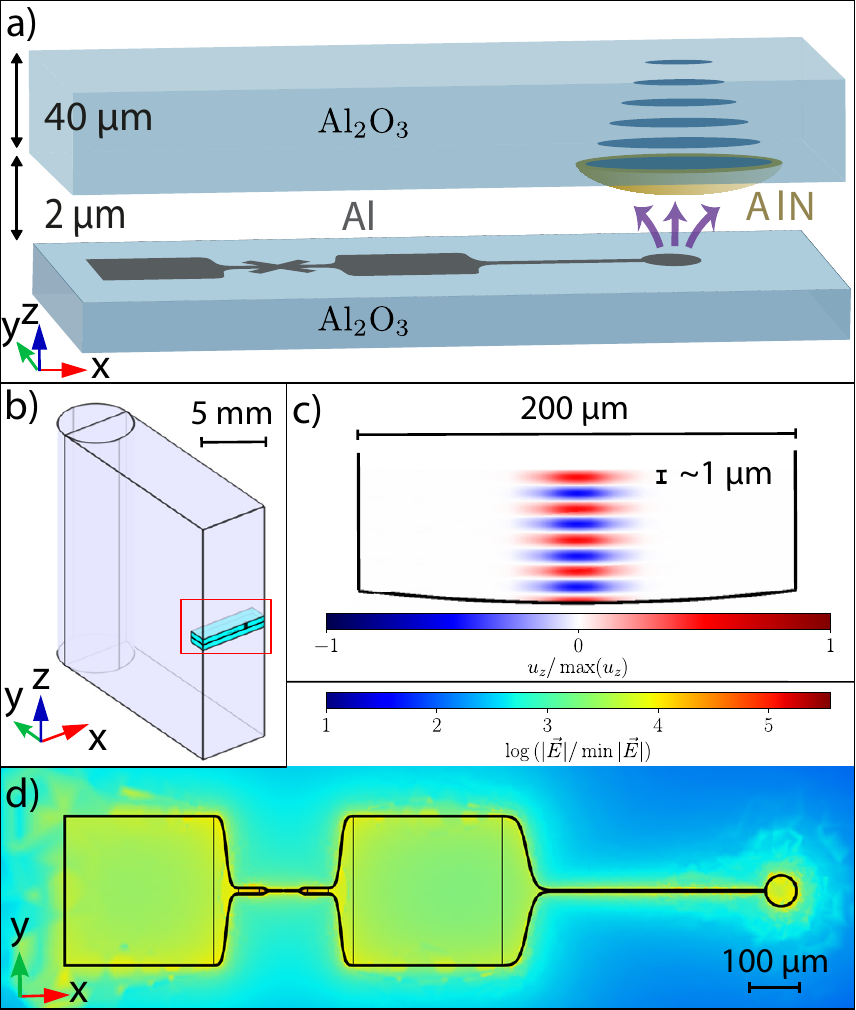}
\caption{\label{fig:device}%
Simulating the $\hbar$BAR. \textbf{a)} A schematic of the $\hbar$BAR device. The bottom chip is a 3-D transmon qubit, modified with an antenna (the extension at the right), on a dielectric substrate. The top chip hosts a HBAR, formed by a piezoelectric dome, positioned above the qubit antenna. The purple arrows represent the electric field of the qubit-like mode. \textbf{b)} Full simulation space defined by half of a rectangular SC microwave cavity with with cylindrical ends (lavender volume). The simulation is symmetric about the $x$-$z$ plane closest to the viewer. The cyan structure in the red rectangle is the $\hbar$BAR. \textbf{c)} Displacement field of a simulated fundamental HBAR mode in a 2-D slice through the symmetry axis. This image is not to scale and is cropped to show the bottom $\text{20}\%$ of the HBAR so that the curvature of the dome and displacement profile of the mode are clearly visible. The diameter of the HBAR and half wavelength of the acoustic mode are indicated by scale bars. \textbf{d)} Electric field of the qubit-like mode centered in a region around the 3-D transmon.}
\end{figure}

We use the method presented in this work to simulate an $\hbar$BAR similar to those used in Refs.~\cite{Chu2018, vonLuepke2021parity}. This device is comprised of two chips: The first has one single-junction transmon qubit with one pad extended to form an antenna, and the second has a piezoelectric dome which transduces the qubit's electric field into mechanical modes of the HBAR (Fig.~\ref{fig:device}a). The two substrates are bonded together such that the antenna is aligned underneath the dome~\cite{vonLuepke2021parity}. Then the assembly is situated in a 3-D microwave cavity (Fig.~\ref{fig:device}b). cQED elements are simulated by applying a perfect electric conductor boundary condition to the superconductors (the surfaces of the microwave cavity and the leads of the qubit) and representing the Josephson junction as a lumped element inductor~\cite{Nigg2012}. In our simulation, the substrate of both the qubit and HBAR chips is c-axis oriented sapphire and the piezoelectric dome is made of c-axis oriented aluminum nitride (AlN).

To perform a 3-D full-wave eigenmode simulation of such a device, several simplifications have to be made. The whole device is symmetric about the $x-z$ plane at $y=0$, so we can use symmetry boundary conditions to reduce the simulation space. We observe that the long-lived modes of the HBAR are confined inside a cylindrical volume with a small transverse area in the $x-y$ plane (Fig.~\ref{fig:device}c). We therefore only simulate the mechanical fields inside a cylindrical volume with the radius of the piezoelectric dome. Finally, we simulate an HBAR with a substrate thickness of 40~$\mu$m, an order of magnitude smaller than the 420~$\mu$m thick devices in Refs.~\cite{Chu2018,vonLuepke2021parity}, in order to speed up the simulations. A detailed description of the device model can be found in Appendix~\ref{sec:Setup}.

\subsection{Extracting Hamiltonian parameters}
An eigenmode simulation returns a set of field distributions which are labeled by their eigenfrequencies. In the unhybridized approach, these are $\ul{E}_n$ and $\ul{u}_m$. We use an overlap integral to extract the coupling rates between the unhybridized electromagnetic and mechanical modes~\cite{yang2005introduction} 
\begin{equation}\label{eq:couplingRates}
    g_{nm} = A_{nm}\int_{V, \text{piezo}} \ul{E}_n \cdot \ul{\ul{\ul{e}}}^T : \ul{\ul{\varepsilon}}_m \dd V
\end{equation}
where $\ul{\ul{\ul{e}}}^T$ is the transpose of the piezoelectric tensor, $\ul{\ul{\varepsilon}}_m$ is the strain tensor derived from $\ul{u}_m$, and the proportionality constant $A_{nm}$ comes from normalizing the fields to that of a single photon and phonon. We justify this formula by deriving the piezoelectric Hamiltonian in a multimode Jaynes-Cummings form in Appendix~\ref{sec:piezoHam}, where Eq.~\ref{eq:gFromUncoupled} is the full expression for $g_{nm}$.

For the hybridized eigenmode simulations, we obtain the electric and displacement fields $\ul{E}_k$ and $\ul{u}_k$, respectively, for each $\xi_k$. COMSOL also computes several derived quantities; in this work, we make use of the electric displacement $\ul{D}_k$, strain $\ul{\ul{\varepsilon}}_k$, and stress $\ul{\ul{S}}_k$ fields, the current through the $j$th junction element $I_{kj}$, and the time-averaged global electrical and strain energies $\overline{\mathcal{E}_{\text{elec},k}}$ and $\overline{\mathcal{E}_{\text{strain},k}}$. In order to use the EPR method, one must calculate the energy-participation ratio of the $k$th mode in the energy of $j$th junction. In Appendix~\ref{sec:EPRandHybrid}, we show that the EPRs of the hybridized qubit-HBAR modes can be written as
\begin{equation}
    p_{kj} = \frac{W_{kj}}{\overline{\mathcal{E}_{\text{elec},k}} + \overline{\mathcal{E}_{\text{strain},k}}}
\end{equation}
where the average inductive energy $W_{kj} = \frac12 L_j I_{kj}^2$ and the time-averaged electrical and mechanical energy stored in the system for mode $k$ $\overline{\mathcal{E}_{\text{elec},k}}$, $\overline{\mathcal{E}_{\text{strain},k}}$ are defined by Eqs.~\ref{eq:avgEEnergy} and~\ref{eq:avgSEnergy} in terms of the fields $\ul{E}_k$ and $\ul{\ul{\varepsilon}}_k$. These quantities are easily obtained from the simulation solutions: $L_j$ is the lumped element inductance that we define, while $I_{kj}$, the electric current through the $j$th lumped element, is calculated from the solution as $I_{kj} = \frac1w\int_{\text{elem } j} \ul{J}_k\cdot \ul{t} \dd S$, where $w$ is the junction width, $\ul{J}_k$ is the surface current, and $\ul{t}$ a unit direction vector. In Appendix~\ref{sec:EPRandHybrid}, we show that the ZPF of the junction's flux can be written in terms of the extracted EPR 
\begin{equation}
    p_{kj} = \frac{E_j\phi_{kj}^2}{\frac{1}{2}\hbar \xi_k},
\end{equation}
which in turn defines the Hamiltonian of Eq.~\ref{eq:BBcoupled}. Furthermore, the EPR can be used to calculate various loss mechanisms that arise from or are modified by the hybridization of the modes, as shown in Appendix~\ref{sec:loss}.

\subsection{Finite element considerations}

In finite element method (FEM) simulations, space is divided into polyhedra whose vertices define a mesh. Building this mesh, or ``meshing," is done automatically in modern FE software, but some user input is almost always needed in the case of more involved geometries. Finding a mesh that is coarse enough so that the simulation runs in a reasonable time but fine enough to capture all the important physical phenomena can be challenging.

A rule of thumb for meshing FE simulations is to use at least five meshing elements per wavelength~\cite{comsol_ref}. We can immediately see the issue for hybridized eigenmode simulations: the software has to simultaneously solve for $\ul{E}$ and $\ul{u}$, which have wavelengths that differ by five orders of magnitude at the same frequency. In the case of $\ul{u}$, for which $\lambda \sim 1~\mu$m in typical GHz frequency cQAD devices, a fine mesh can quickly make the simulation intractable if defined over a too big volume. However, we show that even in the case a HBAR, which has a relatively large volume compared to most mechanical resonators used in cQAD systems, a meshing procedure can be found to keep the simulation at a reasonable size while resolving all of the relevant physics. 

The mesh additionally needs to be optimized to reduce the number of so-called \textit{spurious modes}. These unphysical modes are a source of inaccuracy in many FEM applications~\cite{Corr_SpuriousSolid_1972, Rhaman_SpuriousEM_1984, Winkler_spuriousSM_1984}. A handmade mesh was created in order to minimize the number of elements and spurious solutions (the meshing procedure and parameters are reported in Appendix~\ref{sec:Setup}). Our simulations were able to solve for 150 eigenmodes in under 2 hours on a computer with 64 GB of memory. Our meshing procedure drastically reduced, but could not completely eradicate, spurious modes.

\section{Results}

\subsection{Unhybridized eigenmode approach} \label{sec:resultsUnhybridized}

\begin{figure}[htb]
    \centering
    \includegraphics[width=\linewidth]{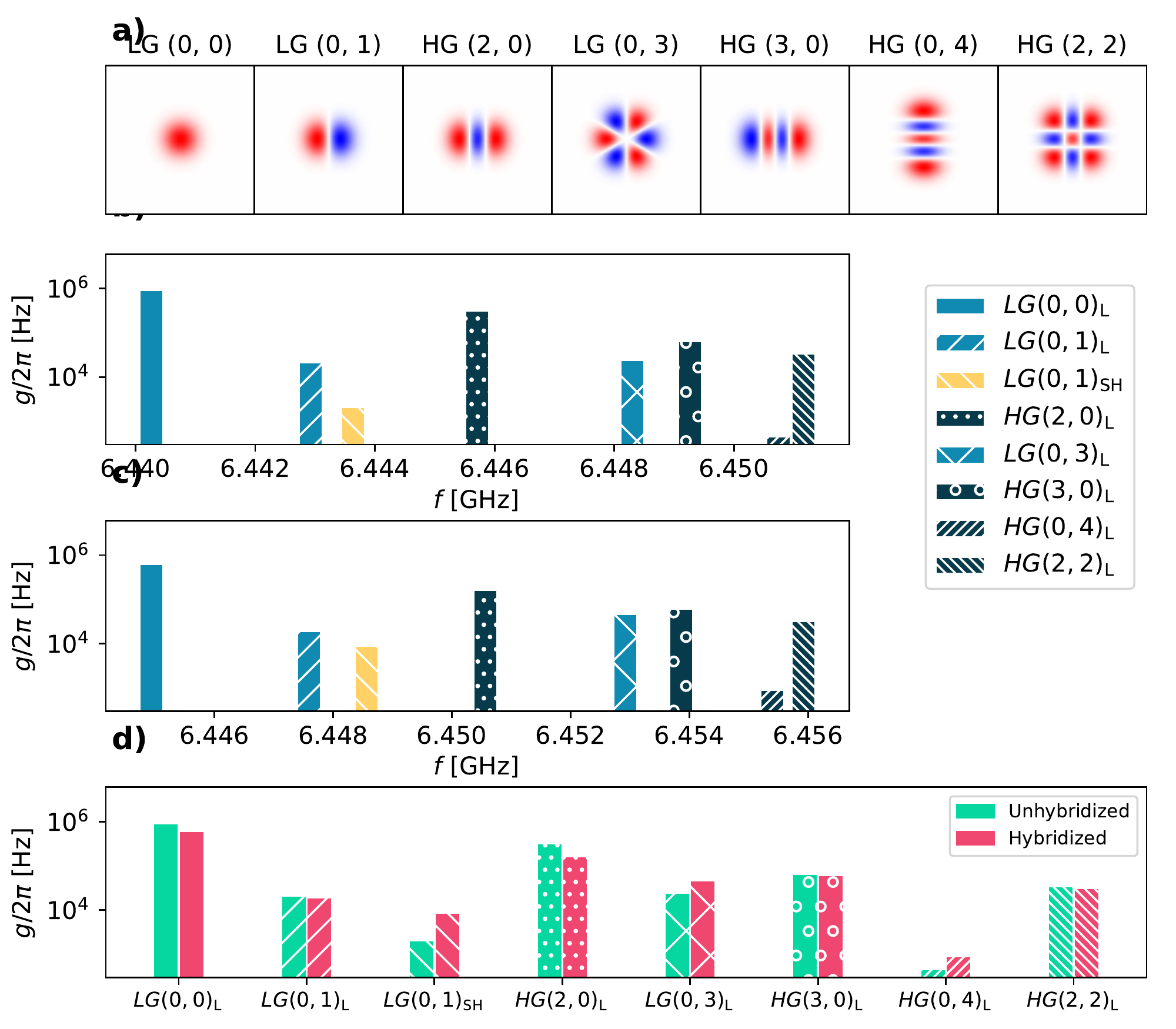}
    \caption{Acoustic modes and electromechanical coupling rates. 
    \textbf{a)} 2-D longitudinal displacement ($u_z$) profiles of acoustic modes observed in our simulations. 
    \textbf{b)} An acoustic spectrum with electromechanical coupling rates calculated using the unhybridized eigenmode approach.
    \textbf{c)} An electro-acoustic spectrum with electromechanical coupling rates calculated using the hybridized eigenmode approach.
    \textbf{d)} A mode-by-mode comparison of the electromechanical coupling rates extracted from unhybridized (green) and hybridized (red) simulations. 
    The coupling rates in \textbf{b)}, \textbf{c)}, and \textbf{d)} refer to a qubit mode at $\omega=2\pi \times 6.424$ GHz.}
    \label{fig:allg_v3}
\end{figure}

We choose to solve the unhybridized solid mechanics eigenmode simulations near a frequency corresponding to $\lambda \approx 1800$ nm in both sapphire and AlN. At this frequency, the mode has half a wavelength in the piezoelectric dome which is expected to maximize the overlap between the acoustic mode's strain field with the piezoelectrically-induced external stress resulting from the qubit mode's electric field. With our geometry, this corresponds to modes with a longitudinal mode number $q=49$. A cross-section of such a mode is shown in Fig.~\ref{fig:device}c. Because the resonator is a 3-D object, it supports many modes with this longitudinal number with different patterns in the transverse plane, as well as different polarizations. We observe both Laguerre-Gaussian (LG) and Hermite-Gaussian (HG) modes in our results. The 2-D profiles of these modes are represented in Fig.~\ref{fig:allg_v3}a. Note that we always show the 2-D profile corresponding to $u_z$ irrespective of the mode's polarization, see also Appendix~\ref{sec:PostProc}. While we are mainly interested in longitudinal-like polarized modes (as defined in Eq.~\ref{eq:pol}), the anisotropy of the material's piezoelectric tensor may give shear-like polarized modes nontrivial coupling to the qubit, making them interesting to study as well. We also perform an unhybridized EM simulation to compute the coupling rates, as described in the next paragraph. An example electric field distribution can be seen in Fig.~\ref{fig:device}d, and an example mechanical displacement field distribution can be seen in Fig.~\ref{fig:device}c in 2-D and Fig.~\ref{fig:spurious_vs_phys}b in 3-D. 

Using these results, overlap integrals are performed between the EM qubit mode ($n=q$ in Eq.~\ref{eq:couplingRates}) and the mechanical modes in order to compute the two-mode coupling rates $g_{qm}$ (hereafter referred to as g for simplicity) and plot them in Fig.~\ref{fig:allg_v3}b against the mechanical modes' frequencies.

We find that the zeroth transverse order mode of a 40 $\mu$m HBAR has a qubit-phonon coupling rate $g \approx 2\pi \times 1$ MHz. The higher-order transverse modes have lower coupling rates due to their overlap mismatch with the qubit's electric field profile. We note that an antenna radius of $r=20~\mu$m was chosen, but it could be optimized in shape to increase the coupling rate to any of the observed acoustic modes.

\subsection{Hybridized eigenmode approach} \label{sec:resultsHybridized}
Next, we turn on the piezoelectric coupling between the \textbf{solid} and \textbf{emw} interfaces. To test our implementation and demonstrate the capabilities of this simulation framework, we perform eigenfrequency simulations while sweeping the qubit's inductance $L$ such that the frequency of the qubit-like mode (defined as the mode with highest EPR) intersects the set of high-overtone HBAR modes mentioned in the previous section. Note that the acoustic mode frequencies shift up by about 5 MHz compared to the uncoupled modes because the piezoelectric effect increases the stiffness of the AlN material~\cite{chen2018electric}. 

\begin{figure}[tb]
    \includegraphics[width=\linewidth]{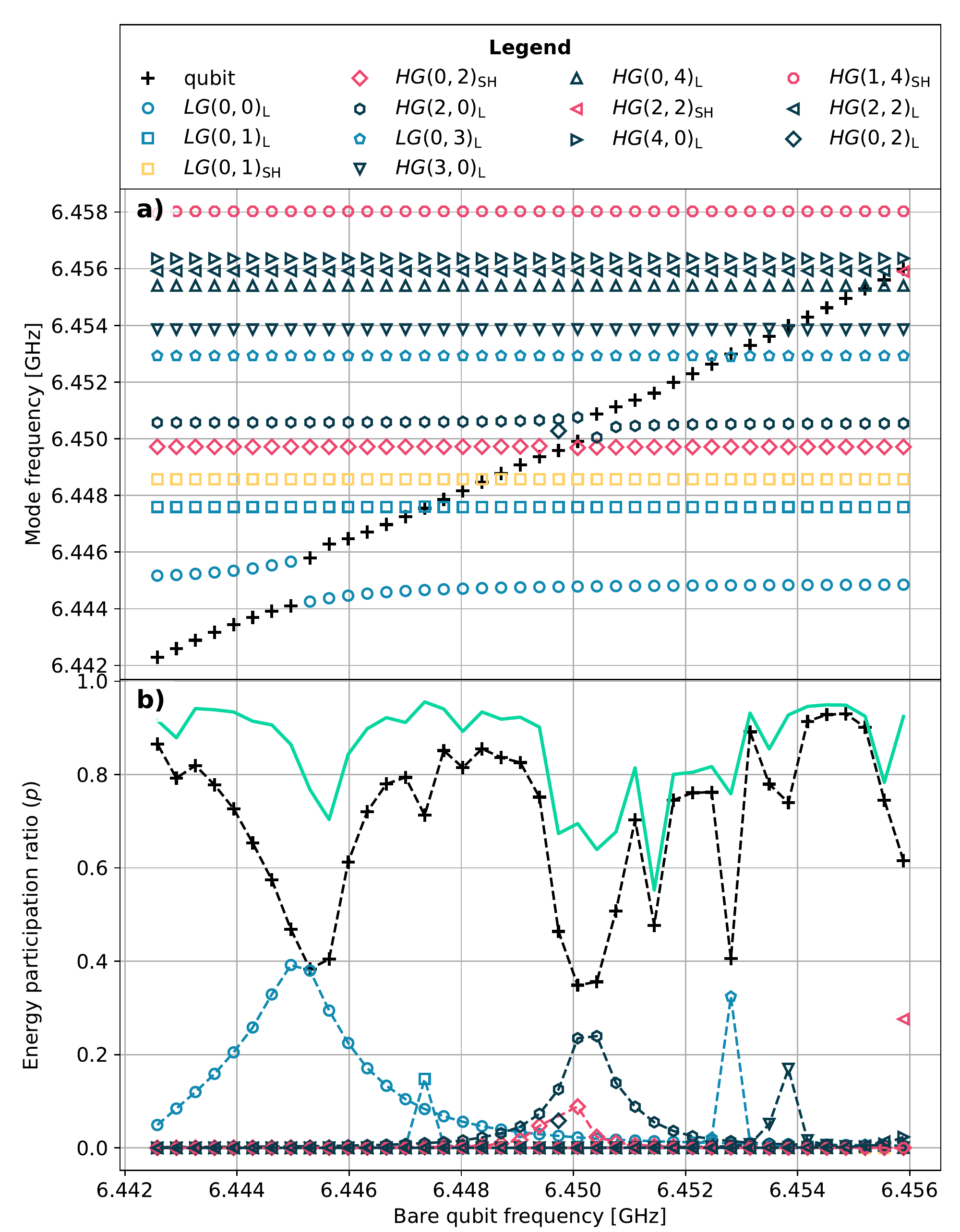}
    \caption{Results of hybridized electromechanical simulations. \textbf{a)} The mode spectrum near the acoustic fundamental longitudinal mode at $\omega=2\pi \times 6.445$ GHz show several avoided crossings. The size of each avoided crossing indicates the coupling between the mechanical mode and the qubit. The legend above this plot applies to both subfigures. \textbf{b)} Energy-participation ratios of each mode shown in Fig.~\ref{fig:fig3}a to the junction. Both longitudinal-like and shear-like modes are labeled according to their $u_z$ profiles. The green line indicates the sum of the EPRs of all of the modes in the figure.}
    \label{fig:fig3}
\end{figure}

The eigenfrequencies found in the simulations are shown in Fig.~\ref{fig:fig3}a and are plotted against the frequency of the unhybridized (UH) qubit mode. We observe avoided crossings in good agreement with the values of $g$ computed in the previous section. This comparison between the unhybridized and hybridized approaches provides a sanity check for the physics modeling. The EPRs calculated from this frequency sweep, shown in Fig.~\ref{fig:fig3}b, are another way to visualize the hybridization of the qubit with the acoustic modes. When the qubit hybridizes with a mechanical mode, a significant fraction of the qubit-like mode EPR is allocated to the mechanical-like mode. 

One challenge of analyzing these simulations is the presence of a significant number of spurious modes in the results, which are a source of inaccuracy in the derived quantities. We find them when looking for GHz-frequency eigenmodes of the displacement field in both unhybridized and hybridized simulations. These modes appear as several point-like defects on mesh edges and nodes (see Appendix~\ref{sec:spur}). We observed that simulations with finer transverse meshing tended to have more spurious modes. 

To address this issue, we identify the so-called "physical modes" in post-processing by finding the modes whose displacement fields best match those of the LG and HG modes. These, together with the qubit mode, are used for further analysis, while the rest are discarded as spurious modes (see Appendix~\ref{sec:Setup}). This process is not ideal, however, since the physical modes can be hybridized with the spurious modes through their coupling to the qubit. This causes an issue which we refer to as EPR dilution, where part of the physical mode's EPR is distributed to spurious modes with nearby frequencies. The effect of EPR dilution is evident for certain bare qubit frequencies in Fig.~\ref{fig:fig3}b where the EPRs of all the modes sum to less than one (eg. around 6.451 GHz). Evidently, however, the effect of EPR dilution is on the 20\% level.

\begin{table}[ht]
    \centering
    \begin{tabular}{|c|c|c|c|}
        \hline
        Mode & Qubit & LG(0,0) & HG(2,0) \\
        \hline
        $f$ [GHz] & 6.424 & 6.445 & 6.451 \\
        \hline
        $p$ & 0.95 & $1.4 \times 10^{-3}$ & $5.4 \times 10^{-5}$ \\
        % \hline
        % $p$ (adjusted) & 0.95 & $1.2 \times 10^{-3}$ & $1.0 \times 10^{-4}$ \\
        \hline \hline
        $\chi_{k, l} / 2\pi$ [Hz] & Qubit & (0,0) & (1,0) \\
        \hline
        Qubit & $5.2\times 10^8$ & $4.4\times 10^4$ & $2.3 \times 10^3$ \\
        \hline
        (0,0) & - & $1.1 \times 10^3$ & $1.7 \times 10^2$ \\
        \hline
        (1,0) & - & - & $1.7\times 10^0$ \\
        \hline
    \end{tabular}
    \caption{Top: Frequencies and EPRs of the qubit-like mode and the two most strongly coupled acoustic modes. Bottom: Pair-wise (cross-) Kerr couplings $\chi_{k, l} / 2\pi$ between the  modes.}
    \label{tab:dispersive}
\end{table}

We now focus on the case of a large detuning between the qubit mode and the same family of LG acoustic modes as in Figs.~\ref{fig:allg_v3}b, \ref{fig:fig3}a and~\ref{fig:fig3}b in order to demonstrate the extraction of design quantities in the dispersive regime. We choose a value for the junction inductance such that the unhybridized qubit is $\Delta^{\text{UH}}_{q, (0,0)} \approx 2 \pi \times 21$ MHz below the LG(0, 0) mode. We compute the self- and cross-Kerr couplings using the EPRs, from which we can further compute the mode's anharmonicity $\alpha_k = \frac12\chi_{k,k}$ and its Lamb shift $\Delta_k = \frac12 \sum_l \chi_{k, l}$. 

The qubit's anharmonicity can be compared with its capacitive energy $E_C = \frac{e^2}{hC}$, since $\alpha_q \approx \frac{E_C}{2h}$~\cite{Blais2021cqed} for transmons. We can infer the capacitance $C$ of the qubit from the slope of its inverse squared frequency using $\omega_q^{-2} = LC$ and obtain $C=67$ fF, finally giving an expected anharmonicity of $\frac{E_C}{2h} = 288$ MHz. This is in reasonably good agreement with the EPR result of $\alpha_q = 261$ MHz (see Table~\ref{tab:dispersive}), considering that $\frac{E_C}{h}$ overestimates $\alpha_q$~\cite{Koch2007}. 

We further verify the results in Table~\ref{tab:dispersive} by using them to compute the coupling rate $g$ between the corresponding unhybridized modes with the approximate relationship~\cite{Koch2007}
\begin{equation} \label{eq:gfromp}
    g_{q,l}^2\approx \Delta_{q,l}^{\text{UH}}\chi_{q,l}\frac{\Delta^{\text{UH}}_{q,l} + \alpha_q}{2\alpha_q}
\end{equation}
The mode numbers $k,l$ either refer to unhybridized modes (on $g$ and $\Delta^{\text{UH}}$) or to their hybridized counterparts (on $\chi$ and $\alpha$). The results are shown in Fig.~\ref{fig:allg_v3}c. Figure~\ref{fig:allg_v3}d shows good agreement between the $g$'s computed using the two methods we described. The discrepancies may be in part explained by EPR dilution and the approximate nature of Eq.~\ref{eq:gfromp}. 

\section{Outlook}
We have demonstrated a technique for the simulation of cQAD devices that unifies electromagnetic and mechanical degrees of freedom under the EPR method. Importantly, we showed that quantum circuits, including SC qubits, can be combined with solid mechanics within the powerful multiphysics framework of COMSOL. By combining the necessary physics interfaces, we can extract the hybridized eigenmodes of the entire system. We showcased this technique by simulating an $\hbar$BAR, and showed that it gives results in good agreement with other methods for estimating device parameters in the dispersive regime. 

The example shown in this work may be the most computationally intensive out of the existing cQAD systems due to the large size of the acoustic resonator. Therefore, applying our methodology to other types of devices should prove relatively straightforward. More generally, our technique may be extended to other types of hybrid quantum systems such as a SC qubit coupled to magnonic resonators~\cite{clerk2020hybrid, lachance2019hybrid} or devices where acoustic resonators interact with other qubits such as color centers~\cite{chen2019engineering} or quantum dots~\cite{wigger2021remote}.

The simulation files used in this paper will be made available in the supplementary material.

\section*{Acknowledgements}
We thank S. Marti, Y. Dahmani, V. Jain, G. Steele, J. Franse, N. Egli, R. Benevides, and U. von Lüpke for valuable discussions. This project has received funding from the European Research Council (ERC) under the European Union’s Horizon 2020 research and innovation programme (Grant agreement No. 948047)

\clearpage

\appendix

\section{Piezoelectric Hamiltonian} \label{sec:piezoHam}

In this appendix, we derive the quantum Hamiltonian for a piezoelectric solid from first principles.

\subsection{Field quantization}

We begin by quantizing the electric and displacement fields. Following the procedure in~\cite{Steck2007quantum}, we will explicitly quantize the displacement field. Following Chapters 3 and 4 of~\cite{royer1999elastic}, we define the displacement field $\ul{u}(\ul{x}, t)$, strain tensor $\ul{\ul{\varepsilon}}(\ul{x}, t) = \frac12\left(\nabla \ul{u}(\ul{x}, t) + \left(\nabla \ul{u}\right)^T(\ul{x}, t)\right)$ and stress tensor $\ul{\ul{S}}(\ul{x}, t)$. In an anisotropic linear material with stiffness tensor $\ul{\ul{\ul{\ul{c}}}}$ and density $\rho$, Newton's second law implies the following wave equation (neglecting body forces)
\begin{equation} \label{eq:elasticwavenobf}
\begin{split}
    \nabla \cdot \ul{\ul{S}} &=\rho \pddiff{\underline{u}}{t} \\
    c_{ijlm} \pddifft{u_l}{x_j}{x_m} &= \rho \pddiff{u_i}{t}
\end{split}
\end{equation}
where step 2 is written in index notation using Einstein's convention, and Hooke's law is used since we assumed a linear elastic material. We split time and space dependencies of a displacement mode assuming a frequency $\Omega$, keeping the convention for naming frequencies from the main text ($\omega$ for electromagnetics and $\Omega$ for solid mechanics)
\begin{equation}
    \ul{u}(\ul{x},t) = u_0 \mathrm{e}^{-i\Omega t} \ul{h}(\ul{x}) + \mathrm{c.c}.
\end{equation}
where $u_0$ is a constant and $\ul{h}(\ul{x})$ is a normalized function that contains the mode shape and polarization of the mode. With $u_0(t) = u_0\mathrm{e}^{-i\Omega t}$, we write the Hamiltonian of the system
\begin{equation}
\begin{split}
    H &= T+V \\
    &= \int \dd V \frac12 \rho \pdiff{u_i}{t}\pdiff{u_i^*}{t} + \int \dd V\frac12 c_{ijlm}\pdiff{u_i}{x_j}\pdiff{u_l}{x_m} \\
    &= \!\begin{multlined}[t]
        \frac12 \rho \Omega^2 \abs{iu_0(t) - iu_0^*(t)}^2 \int \dd V \abs{\ul{h}(\ul{x})}^2 \\+ \frac12\int \dd S n_j c_{ijlm}u_i\pdiff{u_l}{x_m} - \frac12\int \dd V u_i c_{ijlm} \pddifft{u_l}{x_j}{x_m}
    \end{multlined}\\
    &= \frac12 \rho \Omega^2 \abs{iu_0(t) - iu_0^*(t)}^2 +0- \frac{1}{2} \int \dd V u_i \rho \pddiff{u_i}{t}\\
    &= \frac12 \rho \Omega^2 \left(\abs{iu_0(t) - iu_0^*(t)}^2 + \abs{u_0(t) + u_0^*(t)}^2\right) \\
    &= 2 \rho \Omega ^2 \abs{u_0(t)}^2
\end{split}
\end{equation}
where the fourth step assumes no energy leaves the system's volume and uses the wave Eq.~\ref{eq:elasticwavenobf} and the spatial normalization of $\ul{h}$. $n_j$ is the $j$-th element of a unit normal vector to the surface in the second term of step 3. 

We define the following conjugate variables
\begin{equation}
    p = -i\Omega \rho (u_0(t) + \mathrm{c.c}) ~~~~~~ q = (u_0(t)+\mathrm{c.c})
\end{equation}
which gives the following familiar Hamiltonian
\begin{equation}
    H = \frac{p^2}{2\rho} + \frac{1}{2}\rho \Omega^2q^2
\end{equation}
that we quantize using mechanical ladder operators
\begin{equation}
    \hat{p} = -i\sqrt{\frac{\rho \hbar \Omega}{2}}\left(\hat{b}-\hat{b}^\dagger\right) ~~~~~~ \hat{q} = \sqrt{\frac{\hbar}{2\rho \Omega}}\left(\hat{b}+\hat{b}^\dagger\right).
\end{equation}
Identifying $\hat{u}_0(t) = \sqrt{\frac{\hbar}{2\rho \Omega}} \hat{b}$ lets us express the quantum displacement field as
\begin{equation}
    \hat{\ul{u}}(\ul{x},t) = \sqrt{\frac{\hbar}{2\rho \Omega}} \ul{h}(\ul{x})\hat{b}(t) + \text{H.c.}
\end{equation}
and the single mode quantum strain tensor as
\begin{equation}
    \hat{\ul{\ul{\varepsilon}}}(\ul{x},t) = \sqrt{\frac{\hbar}{2\rho \Omega}} \nabla \ul{h}(\ul{x})\hat{b}(t) + \text{H.c.}
\end{equation}
The multimode extension of this derivation is straightforward and also follows the recipe from~\cite{Steck2007quantum}
\begin{align*}
    \hat{\ul{u}}(\ul{x},t) &= \sum_{m=1}^M \sqrt{\frac{\hbar}{2\rho \Omega_m}} \ul{h}_m(\ul{x})\hat{b}_m(t) + \text{H.c.} \\ 
    &= \sum_{m=1}^M \left(\ul{u}_m(\ul{x})\hat{b}_m(t) +\text{H.c.}\right) \\
    \hat{\ul{\ul{\varepsilon}}}(\ul{x},t) &= \sum_{m=1}^M \sqrt{\frac{\hbar}{2\rho\Omega_m}} \nabla \ul{h}_m(\ul{x})\hat{b}_m(t) + \text{H.c.} \\ 
    &= \sum_{m=1}^M \left(\ul{\ul{\varepsilon}}_m \hat{b}_m(t) + \text{H.c.}\right)
\end{align*}
where mode $m$ has frequency $\omega_m$, normalized shape $\ul{h}_m(\ul{x})$ and is created (resp. annihilated) by operator $\hat{b}_m(t)$ (resp. $\hat{b}_m^\dagger(t)$).

The same derivation for the electric field gives
\begin{align*}
    \ul{\hat{E}}(\ul{x},t) &= \sum_{n=1}^N \left(-\sqrt{\frac{\hbar \omega_n}{2\epsilon_0}}\ul{f}_n(\ul{x})\hat{a}_n(t) + \text{H.c.}\right) \\ 
    &= \sum_{n=1}^N \left(\ul{E}_n(\ul{x})\ahat_n(t) + \text{H.c.}\right)
\end{align*}
with $\epsilon_0$ the vacuum permittivity and normalized shape functions $\int_V \ul{\ul{\epsilon_r}}(\ul{x}) \ul{f}^*_n(\ul{x})\ul{f}_m(\ul{x}) \dd V = \delta_{nm}$ where $\ul{\ul{\epsilon_r}}(\ul{x})$ is the relative permittivity in the medium.
\subsection{Piezoelectricity}
Before moving on to the piezoelectric Hamiltonian, we first specify the classical piezoelectric relations we will use.
In a piezoelectric medium, we define the permeability $\ul{\ul{\mu}}$, permittivity at constant strain $\ul{\ul{\epsilon_\varepsilon}}$, density $\rho$, stiffness tensor at constant electric field $\ul{\ul{\ul{\ul{c_E}}}}$, its inverse $\ul{\ul{\ul{\ul{s_E}}}}$ and the strain-charge form piezoelectric coupling tensor $\ul{\ul{\ul{d}}} = \ul{\ul{\ul{\ul{s_E}}}}:\ul{\ul{\ul{e}}}$. We write the piezoelectric constitutive relations in stress-charge from
\begin{equation} \label{eq:piezoConst}
    \begin{pmatrix}
        \ul{\ul{S}} \\ \ul{D}
    \end{pmatrix} =
    \begin{pmatrix}
        \ul{\ul{\ul{\ul{c_E}}}} & -\ul{\ul{\ul{e}}}^T \\
        \ul{\ul{\ul{e}}} & \ul{\ul{\epsilon_\varepsilon}}
    \end{pmatrix}
    \begin{pmatrix}
        \ul{\ul{\varepsilon}} \\
        \ul{E}
    \end{pmatrix}
\end{equation}
In this equation, the product is either a simple matrix product or a double dot product and the sum is always done on the last index(ices). For example,
\begin{align}
    \left(\ul{\ul{\ul{\ul{c_E}}}}\ul{\ul{\varepsilon}}\right)_{ij} \equiv \left(\ul{\ul{\ul{\ul{c_E}}}}:\ul{\ul{\varepsilon}}\right)_{ij} &= \sum_{k,l}c_{E, ijkl}\varepsilon_{kl} \\
    \left(\ul{\ul{\epsilon_\varepsilon}}\ul{E}\right)_i &= \sum_j \epsilon_{\varepsilon, ij}E_j
\end{align}
\subsection{Electromechanical Coupling}
The stored electric and mechanical energies in a system at any time can simply be written as 
\begin{align}
    \mathcal{E}_{\text{elec}}(t) &= \frac12 \int \underline{E}(\ul{x}, t) \cdot \underline{D}^*(\ul{x}, t) \dd V \\
    \mathcal{E}_{\text{strain}}(t) &= \frac12 \int \ul{\ul{S}}(\ul{x}, t) : \ul{\ul{\varepsilon}}^*(\ul{x}, t) \dd V
\end{align}
From these and the piezoelectric constitutive relations, we have that the added energy due to piezoelectricity is given by
\begin{equation} \label{eq:piezoEnergy}
    \mathcal{E}_{\text{piezo}}(t) = -\int \dd V \ul{E}(\ul{x}, t) \cdot \ul{\ul{\ul{e}}}^T : \ul{\ul{\varepsilon}}(\ul{x}, t) 
\end{equation}
which can finally be expanded in terms of the ladder operators
\begin{equation}
    \hat{H}_{\text{piezo}} = -\hbar \sum_{nm} g_{nm}\left(\ahat_n + \adag_n\right)\left(\bhat_m + \bdag_m \right)
\end{equation}
where we have switched to Schrödinger's picture ($\hat{a}(t)\rightarrow \hat{a}$, $\hat{b}(t)\rightarrow \hat{b}$) to remain consistent with the main text and the EPR method, and with
\begin{equation}
    g_{nm} = \frac12 \sqrt{\frac{\omega_n}{\rho \epsilon_0 \Omega_m}} \int \dd V \left(f_n^i\right)^*(\ul{x}) e_{ijk} \left(\nabla \ul{h}_m\right)^{jk}(\ul{x}) \label{eq:gdef}
\end{equation}
In terms quantities that can be easily extracted from simulation results, the coupling rate between the $n$-th electromagnetic and the $m$-th mechanical mode can be expressed as
\begin{equation} \label{eq:gFromUncoupled}
    \frac{g_{nm}}{2\pi} = \dfrac{\sqrt{\frac{\omega_n}{\Omega_m}\frac{1}{\rho \epsilon_0}}\int_{V\text{, piezo}} \ul{E}_n^*(\ul{x}) \cdot \ul{\ul{\ul{e^{T}}}} :\ul{\ul{\varepsilon}}_m(\ul{x}) \dd V}{4\pi \sqrt{\int_{V} \ul{E}_n^{*T}(\ul{x})\ul{\ul{\epsilon_{r}}}\ul{E}_n(\ul{x}) \dd V}\sqrt{\int_{V \text{, SM}} \ul{u}_m^*(\ul{x})\cdot \ul{u}_m(\ul{x}) \dd V}}
\end{equation}

We compute Eq.~\ref{eq:gFromUncoupled} for all solutions of the pure SM simulation using COMSOL's feature \textbf{Volume Integration} in the piezoelectric medium, calling the $E$-field from a chosen solution of the pure EM simulation using COMSOL's operator \texttt{withsol}. For example, one term contributing to the integrand in the numerator of Eq.~\ref{eq:gFromUncoupled} therefore looks like
\begin{equation}
    \verb{conj(withsol('sol2', sext11)) * solid.eXX{
\end{equation}
where \texttt{sol2} refers to the solution of the pure EM simulation, \texttt{sext11} is the 11 component of the external stress $\ul{\ul{S}}_{\text{ext}} = -\ul{\ul{\ul{e^{T}}}} \cdot \ul{E}$ and \texttt{solid.eXX} is the $11$ component of the strain tensor.

\subsection{Electromagnetic-elastomechanical wave equation in a piezoelectric medium} \label{sec:modweq}
We can now derive the full wave equation by using the piezoelectric relations~\ref{eq:piezoConst} along with Maxwell equations, where $\ul{J}$, $\ul{H}$ and $\ul{B} = \ul{\ul{\mu}}\ul{H}$ are the electric current, magnetic and magnetic flux fields, respectively, and $\rho_e$ is the electric charge density
\begin{align}
    \nabla \times \ul{E} &= -i\omega \ul{B} \\
    \nabla \times \ul{H} &= \ul{J} + i\omega \ul{D} \\
    \nabla \cdot \ul{B} &= 0 \\
    \nabla \cdot \ul{D} &= \rho_e,
\end{align}
We combine the two first Maxwell equations together with~\ref{eq:piezoConst} to write a first equation of motion for the electric field
\begin{equation} \label{eq:EOM_PZ1}
    \nabla \times \ul{\ul{\mu}}^{-1} \nabla \times \ul{E}-\omega^{2}\ul{\ul{\epsilon_\varepsilon}}\ul{E} = \omega^{2} \ul{\ul{\ul{e}}}:\ul{\ul{\varepsilon}}
\end{equation}
While combining~\ref{eq:elasticwavenobf} with~\ref{eq:piezoConst} gives
\begin{equation}
    %\nabla \times \ul{\ul{\mu}}^{-1} \nabla \times \ul{E}-\omega^{2}\left(\ul{\ul{\epsilon_\varepsilon}}-\ul{\ul{\ul{e}}} : \ul{\ul{\ul{d^{T}}}}\right) \ul{E}&=\omega^{2} \ul{\ul{\ul{e}}}:\ul{\ul{\varepsilon}} \label{eq:EOM_PZ1} \\
    \nabla \cdot \left(\ul{\ul{\ul{\ul{c_{E}}}}}:\ul{\ul{\varepsilon}}\right)+\rho \omega^{2} \ul{u}=\nabla \cdot\left(\ul{\ul{\ul{e^{T}}}} \cdot \ul{E}\right) \label{eq:EOM_PZ2}
\end{equation}

\section{cQAD dispersive regime considerations} \label{sec:AppDisp}
The form of the Hamiltonian in Eq.~\ref{eq:Hdisp} requires the dispersive regime assumption for all interacting pairs of two electromagnetic modes $(n,n')$; $|\Delta_{nn'}| \gg |\varsigma_{nn'}|$ and for all pairs of an electromagnetic and a mechanical mode $(n,m)$; $|\Delta_{nm}| \gg |g_{nm}|$. It also requires the perturbative assumption
$$\Delta^H_{kl} \gg \frac{E_j}{\hbar}\avg{\hat{\phi}_j}^p~~\forall ~k,l,j \text{ and } \forall~ p\geq 4$$ expressed in the hybridized eigenmode approach (``$H$") , with $\Delta^H_{kl} := \xi_l-\xi_k$.

This condition is usually satisfied in cQED, but not always in cQAD~\cite{vonLuepke2021parity, Arrangoiz2019, Sletten2019}. For example, in the case analyzed in the main text, with a single junction labeled $j=0$, the second assumption is not respected, as $\Delta^H_{q,(0,0)} = 21 \text{ MHz} < \frac{E_0}{\hbar} \phi_{q0}^4 = 310$ MHz where $q$ refers to the qubit mode. This intermediate regime requires an additional transformation to go from the fourth order expansion of Eq.~\ref{eq:BBcoupled} to Eq.~\ref{eq:Hdisp}, namely a Schrieffer-Wolff transformation that removes the term proportional to $\sum_{l\neq q}\cdag_q \chat_q \cdag_l \chat_q + \text{ H.c.}$, which has a time dependence of frequency $\Delta^H_{ql}$ and therefore cannot be neglected in the rotating wave approximation. Using~\cite{Blais2021cqed}, in the simple case of a single acoustic-like mode $l$ with $\phi_{l0} \ll \phi_{q0}$, this changes the expression for the cross-Kerr coupling rates from $\frac{E_0}{\hbar}\phi_{q0}^2\phi_{l0}^2$ to $\frac{E_0}{\hbar}\phi_{q0}^2\phi_{l0}^2\frac{1}{1 + \frac{\alpha_q}{\Delta^H_{ql}}}$. We can see this correction has a significant effect for large $\frac{\alpha_q}{\Delta^H_{ql}}$ ratios, which is the case in our results. In terms of the EPRs, the qubit's anharmonicity does not change, but the cross-Kerr couplings becomes
\begin{equation}
    \chi_{ql} = \frac{\hbar}{E_0} \frac{\xi_q\xi_lp_qp_l}{4}\frac{1}{1+\frac{\hbar}{E_0}\frac{\xi_q^2p_q^2}{8\Delta^H_{ql}}}
\end{equation}

\section{COMSOL simulation setup} \label{sec:Setup}

In COMSOL, all the information needed for a simulation is stored in a single file which, in the software, presents itself as a tree with several levels of nodes. At top level there is the file node, which consists in global definitions, components, studies and results. In a component node, we can define a model's geometry, meshing options and most importantly the physics interfaces, defining which fields will be solved in the model along with their equations of motion. Each of these nodes also feature subnodes which give additional details. In physics interfaces, the subnodes can be domain conditions, such as the main equation of motion or initial values, or boundary conditions like \textbf{fixed} or \textbf{free}.

\subsection{Geometry in detail}

The geometry consists of a $5 \times 30.5 \times 17.8$ mm$^3$ 3-D microwave cavity with two sapphire substrates: a bottom one (the qubit substrate, $5\times 2.6 \times 0.420$ mm$^3$) on top of which the transmon and its antenna sit and a top one (the HBAR substrate, $5\times 2.6 \times 0.04$ mm$^3$) which features the piezoelectric dome, made of AlN, looking down above the antenna. The two substrates are 3.0 $\mu$m apart in the $z$ direction and the piezoelectric dome has a 100 $\mu$m radius and a 900 nm maximum height. 

These are typical values for the devices used in~\cite{ChuScience2017, Chu2018, vonLuepke2021parity}. To keep the simulations light, top substrates of only 40 $\mu$m are used in the simulations used to produce the results of the main text. This approximation is necessary to make the computations tractable, making some numerical results incomparable to actual experiments. They are still useful as proofs of concept and may actually be experimentally relevant in the coming years as one path being explored in the future experiments is the use of thinner HBARs, like the one presented by Blésin \textit{et. al} in a recent proposal for microwave-optical transduction~\cite{blesin2021transduction}. 

As a further simplification, since the whole geometry is symmetric about the $x-z$ plane at $y=0$, the model is cut in half there (axes definitions can be seen in Fig.~\ref{fig:device}) and symmetry boundary conditions are used. 

Since we observed that the elastic waves excited in the HBAR by this configuration where well confined to a small region in the $x-y$ plane, we define a cylinder cut of the HBAR substrate of the same radius as the piezoelectric dome. We will only solve for the displacement field inside of this cylinder instead of the whole HBAR substrate to avoid extending the required fine mesh any more than necessary. The boundaries of this cylinder cut are modeled as low reflecting boundaries to avoid any unphysical reflections. 

\subsection{Physics modeling}

A 3-D microwave cavity is simply empty space surrounded by a perfect electric conductor (PEC). In most FE software, modeling the PEC is done using a boundary condition of the same name, which avoids having to model an actual layer of metal which would need to be meshed and would make the simulation more complex. Losses can be modeled using either scattering boundary conditions or perfectly matched layers (PML) on certain parts of the exterior boundary which correspond to physical objects such as input and output ports. 

The transmon qubit is drawn as a 2-D object. The superconducting aluminum can simply be modeled as a PEC while the Josephson junction is modeled as a linear inductance using a lumped element boundary condition. This recipe without the lumped element can be used to model coplanar waveguides or 2-D resonators.

All the nodes used in the \textbf{Electromagnetic waves} physics interface are detailed here, where an item with a $\blacksquare$ is a \textbf{domain} condition while an item with a $\square$ is a \textbf{boundary} condition:
\begin{itemize}[label=$\blacksquare$]
    \item The \textbf{Wave equation, electric} node is applied to all domains. It defines the EOM for the electromagnetic part of the simulation; the Helmholtz equation in the frequency domain. Without any additional node, this equation reads \\ $\nabla \times \ul{\ul{\mu}}^{-1} \nabla \times \ul{E}-\omega^{2}\ul{\ul{\epsilon}} \ul{E}=0$
    \item The \textbf{effective medium 1} subnode defines a modified relative susceptibility $\ul{\ul{\tilde{\epsilon}_r}} = \ul{\ul{\epsilon_{r, \text{AlN}}}}-\ul{\ul{\ul{e}}} :\ul{\ul{\ul{d^{T}}}}/\epsilon_0$ and applies it to the domain corresponding to the piezoelectric dome. This is the first of three steps for piezoelectric implementation.
    \item The \textbf{External current density} node adds a source term of the form $-i\omega \ul{J}_{\text{ext}}$ to the right-hand-side of the Helmholtz equation for the domain corresponding to the piezoelectric dome. We define $\ul{J}_{\text{ext}} = i\omega \ul{\ul{\ul{e}}}:\ul{\ul{\varepsilon}}$. This is the second step for piezoelectric implementation.
    \item[$\square$] The \textbf{Perfect electric conductor 1 \& 2} nodes define perfectly reflecting boundaries for the electromagnetic fields. The first node is applied to the exterior boundaries (sides of the cavity) while the second one is applied to the transmon's geometry parts since it is implemented as a 2-D object. The boundary equation is simply $\ul{n} \times \ul{E}=0$, where $\ul{n}$ is a normal vector to the boundary element.
    \item[$\square$] The \textbf{Lumped element} node acts like a linear circuit containing at most a resistor, a capacitor and an inductor. It has to be connected to conductors (perfect electric conductors in our case) on two sides. We use it to act like the linear part of our Josephson junction, so we define it as an inductor.
    \item[$\square$] The \textbf{Perfect magnetic conductor} node is simply a symmetry boundary condition for the electric field. This allows us to cut the whole system in half along the $x-z$ plane at $y=0$ since the system is the same on both sides.
\end{itemize}
And for the \textbf{Solid Mechanics} interface, which is only solved for in the piezoelectric dome and a cylinder cut of the HBAR above it, with the same height as the top substrate and the same radius as the dome:
\begin{itemize}[label=$\blacksquare$]
    \item The \textbf{Linear elastic material} node defines the EOM for the solid mechanics part of the simulation in all selected domains (piezoelectric dome and HBAR cylinder cut). The equation is the standard elastic wave equation: $\nabla \cdot \ul{\ul{S}}+\rho \omega^{2} \ul{u}=0$. Without additional nodes we have $\ul{\ul{S}} = \ul{\ul{\ul{\ul{c_{E}}}}}:\ul{\ul{\varepsilon}}$
    \item The subnode \textbf{External Stress} adds the following external stress $\ul{\ul{S}}_{\text{ext}} = -\ul{\ul{\ul{e^{T}}}} \cdot \ul{E}$ to $\ul{\ul{S}}$ in the domain corresponding to the piezoelectric slab. This is the third and final step for piezoelectric implementation.
    \item The \textbf{Prescribed displacement} node can be used to simplify the simulation to only include the $z$ component of the displacement field $u_z$, leaving $u_x$ and $u_y$ at 0 everywhere. This has been shown to be a very good approximation while drastically reducing the number of spurious modes in the results of the simulation.
    \item[$\square$] The \textbf{Free} boundary condition is applied on the sides and bottom of the piezoelectric dome and on top of the HBAR.
    \item[$\square$] The \textbf{Low reflecting boundary} node is used to avoid reflection on the unphysical boundaries on the side of the HBAR's cylinder cut. Note that a perfectly matched layer (PML) is usually preferred to this kind of boundary conditions, but unfortunately in our case it can't be implemented (see note).
    \item[$\square$] The \textbf{Symmetry boundary condition} node is used on the the boundaries on the $x-z$ plane at $y=0$ to cut the system in half as well.
\end{itemize}

\subsection{Meshing procedure}
To properly mesh the HBAR, we need to respect the rule of thumb of 5 elements per wavelength in the longitudinal direction while also resolving higher-transverse-order modes since we expect non-negligible coupling to them. In addition to increasing the simulation size, a finer transverse mesh was also observed to increase the number of spurious modes (see next section). Because no simple metric characterizing the spurious modes is directly available in COMSOL's results, a mesh refinement study could not be used to limit their presence. 

For our model, a handmade mesh was created for the part of the geometry where solid mechanics are solved using a \textbf{mapped} and a \textbf{swept} node, which allow us to control the number of meshing points in all three cylindrical directions using \textbf{distribution} subnodes. For all simulations whose results are reported in this work, the cylindrical region is divided into a shell with inner radius $30 ~\mu$m and a \textbf{mapped} mesh with 10 azimuthal and 6 radial elements and center region, which is a free quad surface mesh with maximum element size $5 ~\mu$m. The rest of the simulation space can then be meshed with automatically generated tetrahedrons, where the only user input are size specifications. We observed that the parameters with the most impact were the maximum element size and $z$-stretching ratio. A light convergence analysis  was performed to ensure the meshing was sufficiently dense near the junction where the electric has a strong gradient. Using this meshing procedure, a hybridized eigenmode simulation finds 150 modes in 2 hours on a computer with 64 GB of memory.

\subsection{Spurious modes} \label{sec:spur}
\begin{figure}[h]
    \centering
    \includegraphics[width=\linewidth]{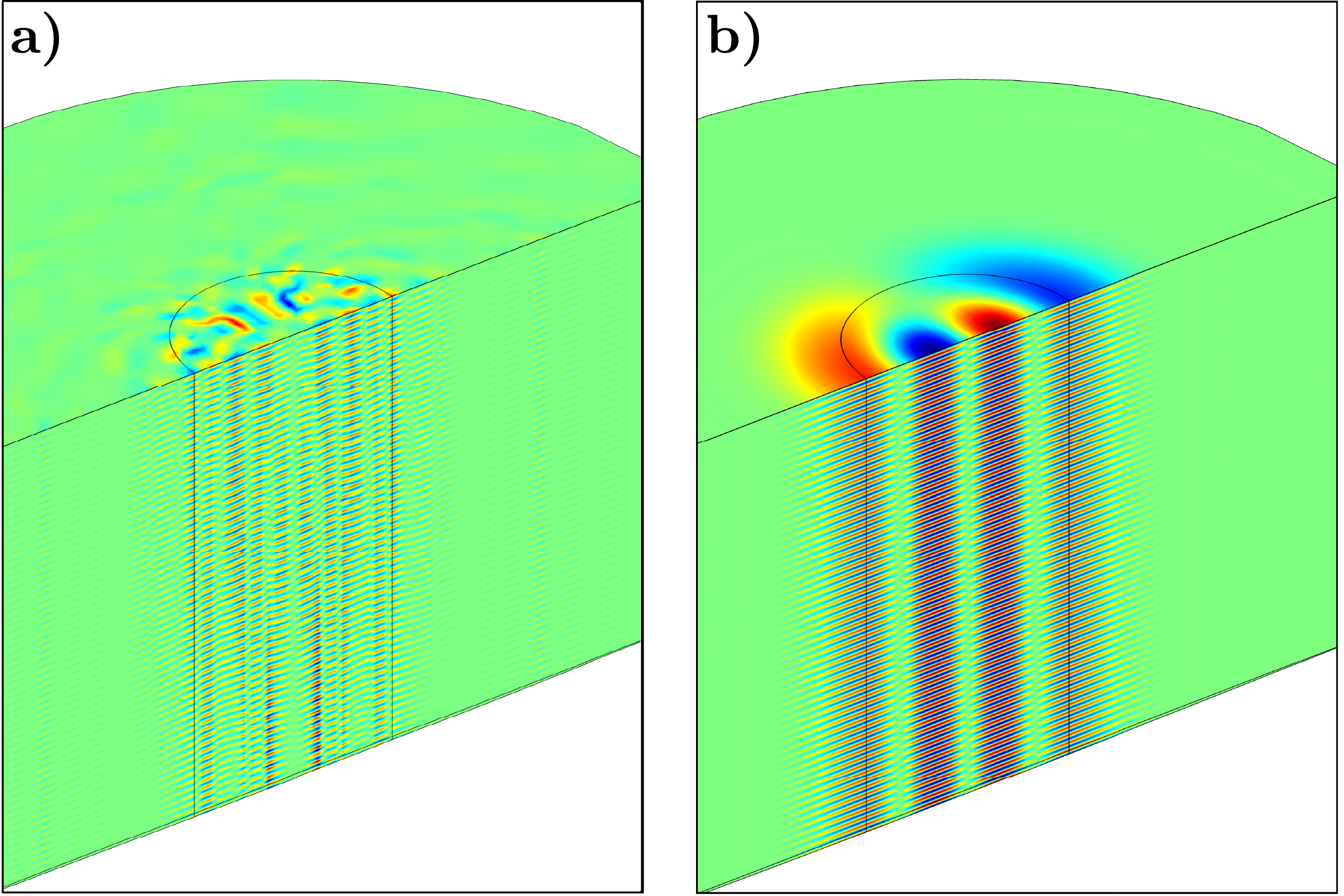}
    \caption{Illustration of two modes with neighboring frequencies in an unhybridized solid mechanics simulation, \textbf{a)} a spurious mode and \textbf{b)} a physical LG(1, 1) mode.}
    \label{fig:spurious_vs_phys}
\end{figure}

FE eigenmode simulations can converge to modes that are not physical, referred to as spurious modes~\cite{Corr_SpuriousSolid_1972, Rhaman_SpuriousEM_1984, Winkler_spuriousSM_1984}. They appear in solid mechanics simulations at GHz frequency and with fine mesh features, yielding field distributions made out of point defects that can be seen in Fig.~\ref{fig:spurious_vs_phys}. In our situation, these modes can appear in greater numbers than physical modes. They cause several problems. First, if one wants to find a certain number of modes  (higher order transverse modes of the HBAR in our case), one typically has to ask the solver for many more modes than this number. Thus, the presence of spurious modes in the results artificially increases the solve time. 

Another detrimental effect of the spurious modes is ``EPR dilution," where the EPR of a physical mode will be shared among several spurious modes that are nearby in frequency. The coupling of the qubit mode to a spurious mode is typically not higher than $5\times 10^3$ kHz, but in certain cases the frequency difference between a spurious mode and a physical or qubit mode can be lower, creating a significant hybridization in the hybrid simulations. This reduces the value of the physical or qubit mode's EPR and the quantities obtained through it, such as the cross-Kerr coupling rate.

No method was found to entirely remove spurious modes from the results of solid mechanics eigenmode simulations of an HBAR. We also could not find any one-number metric that distinguishes them from physical modes, and their estimated convergence error (using COMSOL's error estimates for example) is lower than that of the physical modes, meaning stronger convergence requirements make this issue worse. The only two things one can do to mitigate this problem is optimize the meshing (previous section) and post-process the data.

\subsection{Solver settings and convergence}
All simulations used in this work are done using "eigenmode" COMSOL studies that uses a direct MUMPS solver, with most settings kept as default. However, in the case of hybridized simulations, one change needs to be done in order for the solver to converge to sensible results~\cite{comsol_scaling}. The COMSOL settings "Scaling" and "Residual Scaling" should be set to $10^2$ for the electric field $\vec{E}$ and to $10^{-20}$ for the displacement field $\vec{u}$. This is done in the nodes found under \textbf{Study} $\blacktriangleright$ \textbf{Solver Configurations} $\blacktriangleright$ \textbf{Solution} $\blacktriangleright$ \textbf{Dependant variables}. % We found that more that 20 orders of magnitudes of difference between the fields ($\vec{E}$ scaled by $10^2$ and $\vec{u}$ scaled by $10^{-20}$) worked. % Scaling the fields' residuals by the same amount was found to make all our physical modes well within the convergence requirement for electromagnetics used by \texttt{pyEPR} \cite{pyepr}, the package accompanying the original EPR method \cite{Minev2021a}, which we rewrite here :
%\begin{equation} \label{eq:convergence}
    %\mathcal{C} < 0.1 \text{, with } \mathcal{C} = \dfrac{\abs{E_{\text{cap}} - E_{\text{ind}}}}{\abs{E_{\text{cap}} + E_{\text{ind}}}}
%\end{equation}
%In our case, $E_{\text{cap}}$ needs to be complemented by the piezoelectric energy, given by Eq.~\ref{eq:piezoEnergy}, in order to be properly balanced. Furthermore, as for the original EPR method, the inductive energy has both a magnetic and a lumped-element component :
%\begin{equation*}
    %E_{\text{cap}} = E_{\text{elec}} + E_{\text{piezo}} \text{ and } E_{\text{ind}} = E_{\text{mag}} + E_{\text{junc}}
%\end{equation*}
%Only scaling the fields but not their residuals gives the same spectrum of modes, but with many physical modes having high $\mathcal{C}$, sometimes above 1. While tuning these scaling factors made the simulations work and increased the quality of the modes, it is unfortunately insufficient to remove or even identify spurious modes in the results. While all physical modes were found to have a very low $\mathcal{C}$ (typically $10^{-4}$), spurious modes may have such a low $\mathcal{C}$ as well, sometimes even lower than the best physical mode.

\subsection{Acoustic Polarization and Post-Processing} \label{sec:PostProc}
After the results are computed by COMSOL, we apply a post-processing procedure in order to extract quantities of interest from the simulation. We expect the physical eigenmodes of the HBAR to include Laguerre-Gaussian or Hermite-Gaussian modes with longitudinal (component 33 of $\ul{\ul{\varepsilon}}$) or shear (components 13 or 23) polarization. These modes admit an analytical expression for their longitudinal mode profile ($u_z$ at the top surface). By computing the pointwise distance between the mode profiles of each eigenmode in the results and these analytical mode profiles, we can find the best match in the results, and simply assign all results that aren't good fits for any of the reference modes as spurious modes. Additionally, the longitudinal or shear nature of a physical eigenmode can be simply extracted using, for example, the weight of a tensor component in the strain energy of the mode. Formally, the polarization is attributed to component $c$, with $c$ in \{11, 12, 13, 21, 22, 23, 31, 32, 33 \}, if $c$ respects
\begin{equation} \label{eq:pol}
    \dfrac{\frac14 \text{Re}\int_{V, \text{SM}} S_{c}(\ul{x}) : \varepsilon_{c}^*(\ul{x}) \dd V.}{\overline{\mathcal{E}_{\text{strain},k}}} > \dfrac{\frac14 \text{Re}\int_{V, \text{SM}} S_{\tilde{c}}(\ul{x}) : \varepsilon_{\tilde{c}}^*(\ul{x}) \dd V.}{\overline{\mathcal{E}_{\text{strain},k}}}
\end{equation}
for all other components $\tilde{c}$.
To keep things simple, all physical modes are recognized and labeled according to their $u_z$ profiles. For other mechanical resonator geometries, physical modes can usually also be visually distinguished from spurious ones, but a similar automated method for distinguishing them from spurious modes may need to be developed.

\section{Hybrid EPR method} \label{sec:EPRandHybrid}
The goal of the EPR method, developed by Minev et. al~\cite{Minev2021a}, is to compute the coefficients of the cQED Hamiltonian from the so-called energy-participation ratios. We will mirror a simplified version of the derivation from this paper but in the case of a hybrid Hamiltonian (Eq.~\ref{eq:BBcoupled}). 

The energy of classical mechanical resonator's eigenmode oscillates in time between strain and kinetic energy. Analogously, for a classical electronic circuit, it oscillates between inductive and capacitive energy. In the unhybridized eigenmode approach, for each mechanical (electromagnetic) oscillator in the system, the time-averaged strain (\textit{linear} inductive) energy is equal to the time averaged kinetic (capacitive) energy. Equivalently, each form of energy's time average is equal to half the time-averaged total linear \footnote{The equipartition theorem only applies to quadratic terms of the Hamiltonian} energy for this mode
\begin{equation} \label{eq:Ebalance}
\begin{split}
    \overline{\mathcal{E}_{\text{lin. ind}, n}} = \overline{\mathcal{E}_{\text{elec}, n}} &= \frac12 \overline{\mathcal{E}_{\text{total, EM}, n}} \\
    \overline{\mathcal{E}_{\text{strain}, m}} = \overline{\mathcal{E}_{\text{kin}, m}} &= \frac12 \overline{\mathcal{E}_{\text{total, mech}, m}}.
\end{split}
\end{equation}
$\mathcal{E}_{\text{lin. ind}, n}$ is the sum of the energy stored in the magnetic field $\mathcal{E}_{\text{mag}, n}$ as well as the energies of the lumped element inductances $\sum_j \frac12 L_jI_{nj}^2$. The EPR is then defined as~\cite{Minev2021a} 
\begin{equation}
\label{eq:EPRdef}
    p_{kj} := \frac{\text{linear inductive energy in junction \textit{j} in mode \textit{k}}}{\text{total linear inductive energy in mode \textit{k}}}
\end{equation}

In the hybridized eigenmode approach, the definition of the total linear inductive energy in mode $k$ is extended to include the strain energy:
%Eq.~\ref{eq:Ebalance} applies to the \textit{sum} of strain and linear inductive energies
\begin{equation}
\begin{split}  
     \overline{\mathcal{E}_{\text{lin. ind}, k}}+ \overline{\mathcal{E}_{\text{strain}, k}} &= \overline{\mathcal{E}_{\text{elec},k}} + \overline{\mathcal{E}_{\text{strain},k}}
    \\
    &= \frac{1}{2} \overline{\avg{\hat{H}_{\text{lin}}}}_k = \frac{\hbar}{2}\sum_{l=1}^{N+M} \xi_l\avg{\hat{c}_l^\dagger \hat{c}_l}_k
\end{split}
\end{equation}

We have introduced $\overline{\bullet}$ as the time average and $\avg{\bullet}_k$ as the expectation value of an operator over a state with excitations in a single mode. The numerator of Eq. \ref{eq:EPRdef} is unchanged since a junction's inductive energy does not include any mechanical part. It is defined as the the time-averaged linear inductive \textit{excitation} energy (as opposed to absolute energy) at junction $j$ when only mode $k$ is excited
\begin{equation}
    \overline{\avg{\frac{1}{2}E_j\hat{\phi}_j^2}}_k-\overline{\avg{\frac{1}{2}E_j\hat{\phi}_j^2}}_0
\end{equation}
Defining a general Fock state for this system as 
\begin{equation*}
    \ket{\mu_1, \dots, \mu_{N+M}}
\end{equation*}
we see that writing the EPR in terms of a single-mode Fock state makes it independent of the excitation number and links it to the ZPF of the junction's flux
\begin{equation}
    p_{kj} = \frac{\bra{\mu_k}\frac{1}{2}E_j\hat{\phi}_j^2\ket{\mu_k} - \bra{0}\frac{1}{2}E_j\hat{\phi}_j^2\ket{0}}{\frac{1}{2}\sum_l^{N+M} \hbar \xi_l \bra{\mu_k} \hat{c}_l^\dagger \hat{c}_l\ket{\mu_k}} = \frac{E_j\phi_{kj}^2}{\frac{1}{2}\hbar \xi_k}
\end{equation}
In practice, we extract the denominator of Eq. \ref{eq:EPRdef} by performing finite-sum integrals over the volumes where the fields $\ul{E}$ and $\ul{u}$ are defined
\begin{align}
    \overline{\mathcal{E}_{\text{elec}, k}} &= \frac14 \text{Re}\int_V \underline{E}_k(\ul{x}) \cdot \underline{D}^*_k(\ul{x}) \dd V \label{eq:avgEEnergy} \\
    \overline{\mathcal{E}_{\text{strain},k}} &= \frac14 \text{Re}\int_{V, \text{SM}} \ul{\ul{S}}_k(\ul{x}) : \ul{\ul{\varepsilon}}^*_k(\ul{x}) \dd V \label{eq:avgSEnergy}
\end{align}

\section{Modeling dissipation}\label{sec:loss}

Our simulation framework can also be used to study dissipation in cQAD devices. In this Appendix, we introduce the basics for this next step in the method. The relevant loss mechanisms in a $\hbar$BAR-like device can be separated into two different categories based on how they can be estimated using simulations. In the first case, which we call \textit{semi-analytical loss}, a lossy element (surface or volume) has an intrinsic quality factor that is taken from the literature. Then, its participation in the overall quality factor is weighted by the element's energy-participation ratio which is computed from the results of the simulation. These ratios are referred to as ``lossy" EPRs to distinguish them from the junction EPRs discussed in the rest of the paper, even though the principle is the same. In this section, we illustrate three examples of such dissipation mechanisms: bulk dielectric and surface inductive losses~\cite{Minev2021a, Wang2015, Curtis_PhD_2013} as well as losses due to surface roughness in the acoustic resonator. The second category of losses includes mechanisms that can be fully characterized numerically, so we call it \textit{numerical loss}. One such mechanism present in our system is so-called phonon diffraction loss, where we consider all phonons leaving the center region of the HBAR as lost and quantify this using a numerical flux integration.

\subsubsection{Semi-analytical loss calculations}
For a given EPR-based loss mechanism $\mathcal{L}$, its overall contribution to a mode's quality factor is a weighted inverse sum with contributions from all lossy elements
\begin{equation}
    \frac{1}{Q_k^{\mathcal{L}}} = \sum_l\frac{p^{\mathcal{L}}_{kl}}{Q_l^{\mathcal{L}}}
\end{equation}

\textit{Bulk dielectric losses} of the electromagnetic field are characterized by the loss tangent $\delta_l$ of a lossy solid $l$. This loss tangent is the inverse of an intrinsic quality factor $Q_l^{\text{diel}}$, and the contribution to the overall quality factor of a mode from one such solid is weighted by its energy-participation ratio
\begin{equation}
    p_{kl}^{\text{diel, bulk}} = \frac{1}{\mathcal{E}_k} \frac14\text{Re}\int_{V_l}\ul{E}^*_m \cdot \ul{\ul{\epsilon}}\ul{E}_m \dd V.
\end{equation}
We have defined the total energy of mode $k$ as $\mathcal{E}_k = 2\overline{\mathcal{E}_{\text{elec},k}} + 2\overline{\mathcal{E}_{\text{strain},k}}$ (see Eq.~\ref{eq:Ebalance}).

\textit{Surface inductive losses} are caused by surface currents and result in Ohmic loss. These are characterized by an intrinsic quality factor estimated at unity for metals such as the copper of the microwave cavity, and higher than $10^5$ for SC aluminum~\cite{Minev2021a}. The contribution of a lossy surface $l$ is computed using
\begin{equation}
    p_{kl}^{\text{ind, surf}} = \frac{1}{\mathcal{E}_k} \frac{\lambda_l\mu_l}4 \text{Re} \int_{\text{surf}_l} \ul{H}^*_{k, \parallel} \cdot \ul{H}_{k, \parallel} \dd s
\end{equation}
where $\lambda_l$ is the skin depth of the surface's material and $\mu_l$ its permeability. 

\textit{Acoustic losses due to surface roughness} are estimated using a method from Ref.~\cite{Galliou_lowLossRes_2013}. Surface roughness limits the quality factor to
\begin{equation*}
    Q_{k}^{\text{rough}} = \frac{h^2}{2n_k\sigma^2},
\end{equation*}
where $n_k$ is the longitudinal mode number,  $\sigma^2=\avg{z^2}$ is the height variance of the surface assuming a Gaussian distributed roughness, and $h$ is the height of the HBAR such that $\abs{\ul{q}_k}h=n_k\pi$, where $\ul{q}_k$ is the mode's wave vector. This quality factor only applies to the HBAR, so it has to be weighted by the fraction of energy stored in the mechanics $\frac{2\overline{\mathcal{E}_{\text{strain},k}}}{\mathcal{E}_k}$.
\subsubsection{Numerical loss calculations}
The plano-convex shape of the HBAR was chosen to provide both longitudinal and transverse confinement to the acoustic modes. However, to study the effect of imperfections in this geometry, such as the finite size of the dome, we can use simulations to calculate the acoustic energy leaving the Fabry-Pérot cavity (the region of the sapphire substrate above the piezoelectric dome). This can be treated as loss because, even if the substrate has a finite size and reflecting boundaries, the timescale on which the energy is reflected back into the mode region is much longer than the typical timescale of operations we're interested in~\cite{Chu2018, vonLuepke2021parity}. We compute a quality factor due to \textit{diffraction loss} using a flux integral of the mechanical Poynting vector $\ul{P}_a$
\begin{equation}
    Q^{\text{diff}}_k = \omega_k \frac{\mathcal{E}_k}{\int_{S_d} \ul{P}_a \cdot \dd \ul{\sigma}}
\end{equation}.

Here $S_d$ is a cylindrical surface defines the boundary of the acoustic cavity, and is parametrized by $(x_0 + R\cos \theta, R\sin \theta, z)$, where $x_0=1$ mm is the position of the center of the antenna, $R=90$ $\mu$m, $\theta \in [-\pi/2, \pi/2]$ and $z\in [-0.9, 40]$ $\mu$m, which includes the entire height of the substrate.

\subsubsection{Effects of hybridization on losses}
\begin{figure}[h!]
    \centering
    \includegraphics[width=\columnwidth]{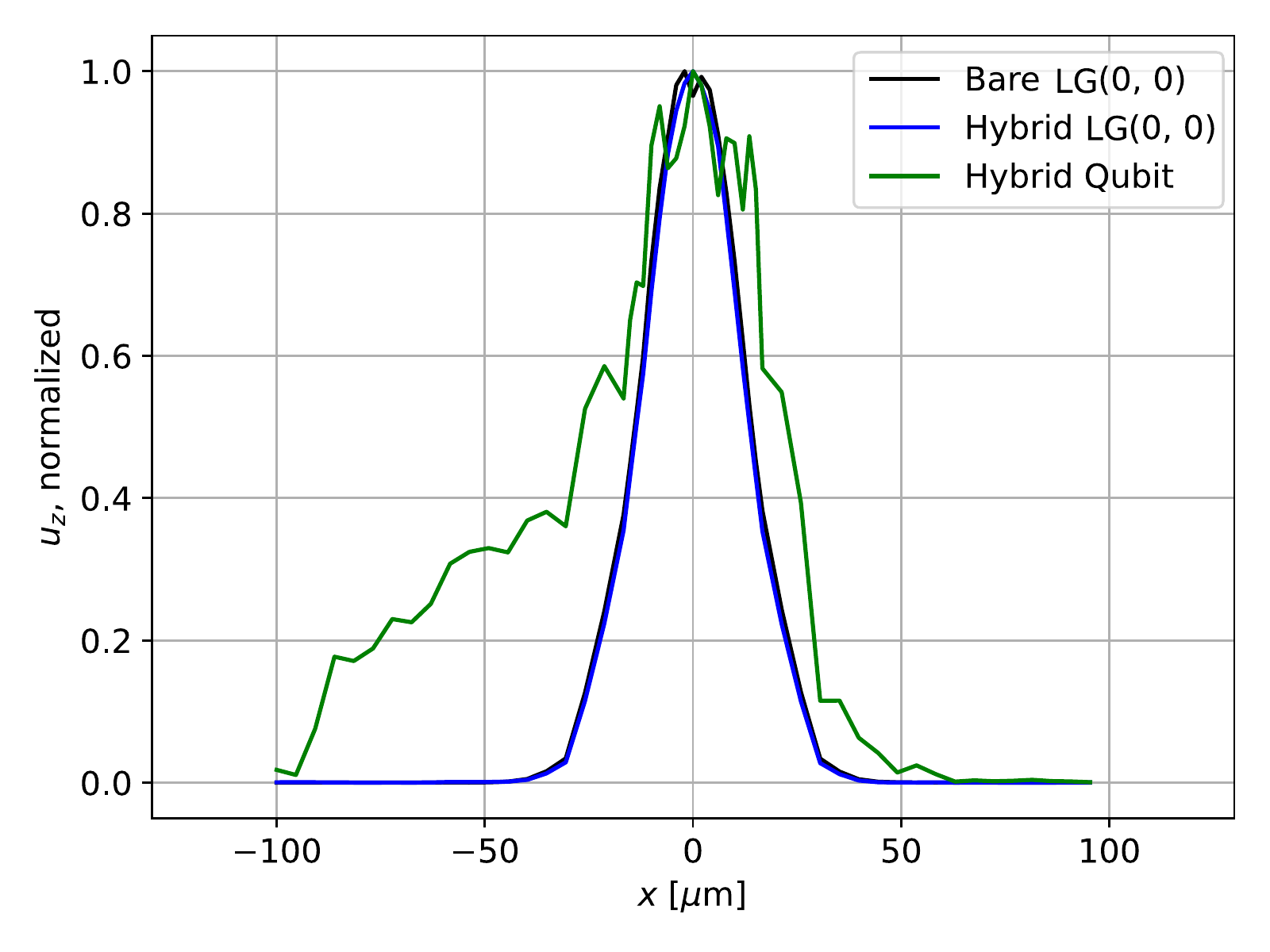}
    \caption{Acoustic displacement along a line defined by the interface between the piezoelectric dome and the sapphire and the symmetry axis of the simulation ($y=0$). The sharp features on the qubit mode are the result of hybridization with spurious modes.}
    \label{fig:profiles}
\end{figure}
An interesting new feature that arises from our simulation framework is the ability to study mechanical losses in hybrid qubit- or cavity-like modes, and electromagnetic losses in mechanical-like modes. These new effects can only be studied once one has access to the full dynamics of the hybridized eigenmodes and are thus a unique feature of the hybridized approach. 

As an example, we show how the qubit mode, once hybridized with the HBAR in the same dispersive regime as in the main text, acquires a new loss channel through phonon diffraction.
Fig.~\ref{fig:profiles} shows the LG(0, 0) mode of the unhybridized and hybridized simulations as well as the qubit-like mode in the hybridized simulation. The displacement profile of the bare mechanical mode (black) and the hybridized mechanical-like mode (green) are almost identical. However, we see that for the qubit-like mode (blue), the piezoelectric coupling to the qubit electric field, which is asymmetric due to the thin lead of the antenna, results in an asymmetric displacement field. This asymmetry is not captured in the unhybridized approach. Such a modification of the acoustic mode shape could lead to additional loss through imperfect mode confinement. Studying these effects using the techniques described in the previous section will be the subject of future work.
\clearpage
\bibliography{COMSOL_sim}% Produces the bibliography via BibTeX.

\end{document}